\documentclass[12pt,a4paper]{article}  
\usepackage{amsmath}
\usepackage{amssymb}
\usepackage{epsfig}
\usepackage{epstopdf}
\usepackage{graphicx}
\usepackage{subfigure}
\usepackage{commath}
\usepackage{float}
\usepackage[left=2.0cm,right=2.0cm, top=2.5cm,bottom=2.5cm]{geometry}
\usepackage{cite}
\usepackage{times}  
\usepackage[colorlinks=true, linkcolor=blue, citecolor=red, urlcolor=blue]{hyperref} 

\begin{document} 
\newcommand{\sheptitle}
{Predictions of effective Majorana neutrino mass under radiative corrections to $\mu-\tau$ reflection symmetry}
\newcommand{\shepauthor}
{ Prokash Pegu\footnote{ E-mail: peguprokash202@gmail.com} and Chandan Duarah \footnote{ E-mail: chandanduarah@dibru.ac.in}}
\newcommand{\shepaddress}
   { Department of Physics, Dibrugarh University,
               Dibrugarh - 786004, India }
\newcommand{\shepabstract} {The search for neutrinoless double beta decay ($0\nu\beta\beta$) is currently one of the key objectives in neutrino physics research. The decay rate of $0\nu\beta\beta$ decay depends on the effective Majorana neutrino mass  $|\langle m \rangle_{ee}|$. In this work we study the numerical prediction of $|\langle m \rangle_{ee}|$ in the scenario of deviation from the $\mu$-$\tau$ reflection symmetry due to radiative corrections, as an extension of our earlier work \cite{pegu}. In \cite{pegu}, we consider an exact $\mu$-$\tau$ reflection symmetry in the light effective Majorana neutrino mass matrix and in the corresponding lepton mixing matrix as well at the seesaw scale. We choose numerical values of all the mixing parameters and neutrino mass eigenvalues at the seesaw scale as inputs and estimate the values of mass eigenvalues and mixing parameters at the electroweak scale due to radiative corrections. We find these low energy predictions consistent with global $3\sigma$ oscillation data. In the present work, we compute the effective Majorana neutrino mass $|\langle m \rangle_{ee}|$ using these low energy values at the electroweak scale. We find that the low energy predictions of $|\langle m \rangle_{ee}|$ are consistent with the latest upper bound $|\langle m \rangle_{ee}|<(0.028-0.122)\ eV$ provided by KamLAND-Zen Collaboration.\\

Key-words: Lepton mixing, $\mu-\tau$ reflection symmetry, radiative corrections, MSSM, Neutrinoless double beta decay ($0\nu\beta\beta$).\\
 }
\begin{titlepage}
\begin{flushright}
\end{flushright}
\begin{center}
{\large{\bf\sheptitle}}
\bigskip\\
\shepauthor
\\
\mbox{}\\
{\it\shepaddress}\\
\vspace{.5in}
{\bf Abstract }
\bigskip
\end{center}
\setcounter{page}{0}
\shepabstract
\end{titlepage}
\section{Introduction}
\indent~ The search for neutrinoless double beta decay ($0\nu\beta\beta$) is one of the major research interests in neutrino physics. 
If this process is ever observed, it will show that lepton number is not conserved and that neutrinos are Majorana particles. The concept of $0\nu\beta\beta$ was first proposed by W. H. Furry \cite{origin}. It is basically a type of $\beta$-decay in which a nucleus with atomic number Z and mass number A transforms into a nucleus with atomic number $Z+2$ and the same mass number, emitting two electrons but no antineutrino:
\begin{equation}
 ^AX_Z\rightarrow\  ^AX_{Z+2} + 2e^-. \label{eq:1}
\end{equation}
From the above reaction, it is clear that the lepton number conservation is violated by two units ($\Delta L=2$). The importance of neutrinoless double beta decay process from both theoretical and experimental points of view has been reviewed in the papers \cite{dell, MJ}. The studies of its phenomenological implications in neutrino physics has been treated considerably in the literature \cite{HB, pet, Nath, chinese, ZY, sarma, Ptecov}. Over the past decade, several experiments such as GERDA \cite{GERDA1}, MAJORANA DEMONSTRATOR \cite{MD}, CUORE \cite{CUORE}, EXO-$200$ \cite{Exo} and KamLAND-Zen \cite{Kam} have searched for this decay in different radioactive isotopes. So far, no signal has been seen, however these experiments have placed very strong lower limits on the $0\nu \beta \beta$ half-life ($T_{1/2}^{0\nu}$). For example, GERDA and MAJORANA DEMONSTRATOR (source isotope $^{76}Ge$) have set $T_{1/2}^{0\nu}> 1.8 \times 10^{26}$ yr and $T_{1/2}^{0\nu}> 8.3 \times 10^{25}$ yr respectively  \cite{GERDA1, MD}. CUORE (source isotope  $^{130}Te$ ) has obtained $T_{1/2}^{0\nu}> 2.2 \times 10^{25}$ yr \cite{CUORE}. Similarly, EXO-$200$ and KamLAND-Zen (source isotope $^{136}Xe$) set limits of $T_{1/2}^{0\nu}> 3.5 \times 10^{25}$ yr and $T_{1/2}^{0\nu}> 3.8 \times 10^{26}$ yr respectively \cite{Exo, Kam}. These limits, in turn, impose upper bounds on the effective Majorana mass $|\langle m \rangle_{ee}|$, which currently lies roughly in the range of a few tens of electron-volts (or meV), depending on the nuclear matrix element calculations used. We note that among these experiments, the most stringent upper bound on $|\langle m \rangle_{ee}|$ is currently provided by the KamLAND-Zen Collaboration, which reports  $|\langle m \rangle_{ee}|<(0.028-0.122)\ eV$ \cite{Kam}. \\
\indent~ The corresponding decay width of the reaction in Eq.(\ref{eq:1}) is given by 
\begin{equation}
    \Gamma^{0\nu}=\left(T_{1/2}^{0\nu}\right)^{-1}=G_{0\nu}| M_{0\nu}(A,\ Z) |^2 |\langle m \rangle_{ee}|^2,\label{eq:2}
\end{equation}
where $T_{1/2}^{0\nu}$ represents the half-life of the decay process, $G_{0\nu}$ stands for the two-body phase-space factor, $M_{0\nu}$ is the nuclear matrix element (NME) and $|\langle m \rangle_{ee}|$ represents the effective Majorana neutrino mass. The expression of $|\langle m \rangle_{ee}|$ is given by 
\begin{equation}
  |\langle m \rangle_{ee}|=  \vert m_{ee} \vert=\vert \sum_{i=1}^3 m_i U_{ei}^2 \vert,\label{eq:3} 
\end{equation}
where $m_i$ are the neutrino mass eigenvalues and $U_{ei}$ are the first row elements of the lepton mixing matrix, also known as the PMNS mixing matrix. In the standard parametrization, the lepton mixing matrix can be parameterized in terms of three mixing angles and three CP violating phases and is given by
\begin{equation}
    U=VP_1,\label{eq:4}
\end{equation} 
where V is given by 
\begin{equation}
  V= \left(\begin{array}{ccc}
c_{12}c_{13}&s_{12}c_{13}&s_{13}{e}^{-i\delta}\\
-s_{12}c_{23}-c_{12}s_{23}s_{13}{e}^{i\delta}&c_{12}c_{23}-s_{12}s_{23}s_{13}{e}^{i\delta}&s_{23}c_{13}\\
s_{12}s_{23}-c_{12}c_{23}s_{13}{e}^{i\delta}&-c_{12}s_{23}-s_{12}c_{23}s_{13}{e}^{i\delta}&c_{23}c_{13}
\end{array}\right),\label{eq:04}
\end{equation}
with $s_{ij}=\sin\theta_{ij}$, $c_{ij}=\cos\theta_{ij}$ ($ij=12,23,13$). Here $\theta_{12}$, $\theta_{23}$, $\theta_{13}$ are the solar mixing angle, atmospheric mixing angle and reactor mixing angle respectively and $\delta$ is the Dirac CP violating phase. Further in Eq.(\ref{eq:4}), $P_1=Diag\left(e^{i\alpha}, e^{i\beta},1\right)$ is the diagonal matrix which contains two Majorana phases $\alpha$ and $\beta$.
With the lepton mixing matrix defined in Eq.(\ref{eq:4}), the expression for the effective Majorana neutrino mass $| m_{ee}|$ becomes
\begin{equation}
    \vert m_{ee}\vert=\vert m_1 c_{12}^2 c_{13}^2 e^{2i\alpha} + m_2 s_{12}^2 c_{13}^2 e^{2i\beta} + m_3 s_{13}^2 e^{-2i\delta}\vert.\label{eq:5}
\end{equation}
\indent~ In this work, we numerically estimate the value of the effective Majorana neutrino mass $|m_{ee}|$ at the electroweak scale $\Lambda_{EW}$, as an extension of our previous work \cite{pegu}. In \cite{pegu}, we assume $\mu-\tau$ reflection symmetry to be preserved at a flavour symmetry scale $\Lambda_{\mu\tau}$ and consider its breaking due to the RG running from $\Lambda_{\mu\tau}$ to  $\Lambda_{EW}$. The flavour symmetry scale $\Lambda_{\mu\tau}$ is chosen at the seesaw scale $10^{14}\ GeV$, while the electroweak scale $\Lambda_{EW}$ is chosen at the top-quark mass scale $m_t=172.76\ GeV$. Using the integral solution method for the one-loop RGE of the effective Majorana neutrino mass matrix $M_\nu$, we derive analytical expressions for the low-energy neutrino mass eigenvalues ($m_1,\ m_2,\ m_3$), mixing angles ($\theta_{12},\ \theta_{23},\ \theta_{13}$) and CP phases ($\delta,\ \alpha$,\ $\beta$) in terms of corresponding high energy parameters $m_1^{\mu\tau}$,\ $m_2^{\mu\tau}$, $m_3^{\mu\tau}$,\  $\theta_{12}^{\mu\tau}$,\  $\theta_{23}^{\mu\tau}$, $\theta_{13}^{\mu\tau}$,\  $\delta^{\mu\tau}$, \ $\alpha^{\mu\tau}$ and $\beta^{\mu\tau}$. Then using these analytical expressions, we estimate the values of all the parameters at $\Lambda_{EW}$, assuming the values of the parameters at the high energy scale as inputs. This work is done within the Minimal Supersymmetric Standard Model (MSSM) framework. In order to see the effects of variation of corresponding predictions at $\Lambda_{EW}$, we consider three different SUSY breaking scales at $\Lambda_s= 1 \ TeV$, $7 \ TeV$ and $14 \ TeV$. Since the $\mu$–$\tau$ reflection symmetry is assumed to be preserved at $\Lambda_{\mu\tau}$, therefore the input values of the mixing angle $\theta_{23}^{\mu\tau}$, the Dirac CP phase $\delta^{\mu\tau}$ and the two Majorana phases $\alpha^{\mu\tau}$ and $\beta^{\mu\tau}$ are constrained by the symmetry. The remaining parameters—three mass eigenvalues ($m_1^{\mu\tau}$, $m_2^{\mu\tau}$, $m_3^{\mu\tau}$) and two mixing angles ($\theta_{12}^{\mu\tau}$, $\theta_{13}^{\mu\tau}$) are treated as free parameters. The input values of these free parameters were chosen such that the resulting low-energy predictions at $\Lambda_{EW}$ are consistent with the global analysis data. With the appropriate choice of the high energy input values of  $m_1^{\mu\tau}$,\ $m_2^{\mu\tau}$, $m_3^{\mu\tau}$,\  $\theta_{12}^{\mu\tau}$,\  $\theta_{23}^{\mu\tau}$, $\theta_{13}^{\mu\tau}$,\  $\delta^{\mu\tau}$, \ $\alpha^{\mu\tau}$ and $\beta^{\mu\tau}$ we finally estimate the values of low energy parameters ($m_1,\ m_2,\ m_3, \theta_{12},\ \theta_{23},\ \theta_{13},\ \delta,\ \alpha$ and $\beta$). With these low energy values of all the lepton mixing parameters and neutrino mass eigenvalues, we estimate the values of $| m_{ee}|$ in the present work. We then check the consistency of these predictions with the current experimental upper bound.  We observe that the low energy predictions of $|m_{ee}|$ for $\Lambda_s = 1\ TeV$, $7\ TeV$ and $14\ TeV$ with $\tan\beta = 30$ and $58$ are consistent with the current upper bound $|m_{ee}| < (0.028 - 0.122)\,eV$ provided by the KamLAND-Zen Collaboration.\\
\indent~ The rest of the paper is organized as follows: in section 2 we present an overview of $\mu-\tau$ reflection symmetry and discuss its deviation induced by radiative corrections. In section 3, we carry out numerical analysis to estimate the low energy values of  $|m_{ee}|$. Finally section 4 is devoted to summary and discussion.
\section{Deviation from $\mu-\tau$ reflection symmetry under radiative corrections}
\indent~ The concept of $\mu-\tau$ reflection symmetry was first introduced by Harrison and Scott \cite{ref1}. They  proposed a specific type of mixing matrix, given by 
\begin{equation}
U_{HS}=\left(\begin{array}{ccc}
u_1&u_2&u_3\\
v_1&v_2&v_3\\
v_1^*&v_2^*&v_3^*
\end{array}\right)
,\label{eq:a}
\end{equation}
where $u_i$'s are real $v_i$'s complex. This mixing matrix is invariant under a combined operation of interchanging $\mu$ and $\tau$ rows and taking the complex conjugate of the entire matrix. This symmetry operation is referred to as 
$\mu-\tau$ reflection operation and the symmetry of the mixing matrix is called $\mu-\tau$ reflection symmetry. This symmetry should also be contained in the mass matrix.
The corresponding mass matrix that respects this symmetry is given by 
\begin{equation}
M_\nu^{\mu\tau} = \left(\begin{array}{ccc}
a^{\mu\tau} & b^{\mu\tau} & \left( b^{\mu\tau}\right)^*\\
b^{\mu\tau} & c^{\mu\tau} & d^{\mu\tau}\\
\left( b^{\mu\tau}\right)^* & d^{\mu\tau} & \left( c^{\mu\tau}\right)^*
\end{array}\right),\label{eq:10}
\end{equation}
where the elements $a^{\mu\tau}$ and $d^{\mu\tau}$ are real \cite{ref2}.\\
\indent~ The significant feature of reflection symmetry is that it corresponds to maximal atmospheric mixing  $ \theta_{23}^{\mu\tau}= \pi/4$ and maximal value of Dirac CP phase $\delta^{\mu\tau}= (\pi/2)/(3\pi/2)$, leaving $ \theta_{13}^{\mu\tau}$ and $ \theta_{12}^{\mu\tau}$ behind arbitrary. However, the reflection symmetry is not shown off in the mixing matrix given in Eq.(\ref{eq:04}), while the maximal values of $ \theta_{23}^{\mu\tau}$ and $\delta^{\mu\tau}$ are substituted, which has been discussed in Ref.\cite{duarah}. Basically, for $ \theta_{23}^{\mu\tau}= \pi/4$ and $\delta^{\mu\tau}= (\pi/2)/(3\pi/2)$, the elements of the  first row of the resulting  mixing matrix are not real, which violates the reflection symmetric nature of $U_{HS}$.
Therefore, in order to preserve reflection symmetry, we consider a complete parametrization of the lepton mixing matrix including the three unphysical phases. Accordingly, the mixing matrix is parametrized as
\begin{equation}
    U=P_2V P_1, \label{eq:b}
\end{equation}
where $P_2=Diag \left( e^{i \phi_1} , e^{i \phi_2}, e^{i \phi_3}\right)$ is the diagonal matrix which contain the three un-physical phases $\phi_1$, $\phi_2$, $\phi_3$, while $V$ and $P_1$ are defined in Eq.(\ref{eq:4}) and (\ref{eq:04}). Since the reflection symmetry allows two possible values of $\delta^{\mu\tau}$, accordingly we consider two cases:
\begin{itemize}
\item \textbf{Case-I:} \ $ \theta_{23}^{\mu\tau}= \pi/4$ and $\delta^{\mu\tau}= \pi/2$,
\item \textbf{Case-II:}\ $\theta_{23}^{\mu\tau}= \pi/4$ and $\delta^{\mu\tau}= 3\pi/2$.
\end{itemize}
Then in order to preserve reflection symmetry, the values of Majorana phases and unphysical phases are fixed as \cite{duarah}  
  \begin{equation}
  \alpha^{\mu\tau}=\beta^{\mu\tau}= \frac{3\pi}{2}, \ \ \phi_1^{\mu\tau}= \frac{\pi}{2},\ \  \phi_2^{\mu\tau}= \phi_3^{\mu\tau}= 0 \label{eq:7}
  \end{equation}
 for Case-I and 
 \begin{equation}
  \alpha^{\mu\tau}=\beta^{\mu\tau}= \frac{\pi}{2}, \ \ \phi_1^{\mu\tau}= \frac{3\pi}{2},\ \ \phi_2^{\mu\tau}= \phi_3^{\mu\tau}=0  \label{eq:8}
  \end{equation} 
for Case-II. With $ \theta_{23}^{\mu\tau}= \pi/4$, $\delta^{\mu\tau}= (\pi/2)/(3\pi/2)$ and the above choices of CP phases
the mixing matrix in Eq.(\ref{eq:b}) takes the form
 \begin{equation}
U^{\mu\tau}= \left(\begin{array}{ccc}
s_{12}^{\mu\tau} c_{13}^{\mu\tau} &s_{12}^{\mu\tau} c_{13}^{\mu\tau} &s_{13}^{\mu\tau}\\
\frac{1}{\sqrt{2}}( c_{12}^{\mu\tau} s_{13}^{\mu\tau} \pm is_{12}^{\mu\tau})&\frac{1}{\sqrt{2}}(s_{12}^{\mu\tau}s_{13}^{\mu\tau}\mp ic_{12}^{\mu\tau})&\frac{1}{\sqrt{2}}c_{13}^{\mu\tau}\\
\frac{1}{\sqrt{2}}(c_{12}^{\mu\tau}s_{13}^{\mu\tau} \mp i s_{12}^{\mu\tau})&\frac{1}{\sqrt{2}}(s_{12}^{\mu\tau}s_{13}^{\mu\tau}\pm i c_{12}^{\mu\tau})&\frac{1}{\sqrt{2}}c_{13}^{\mu\tau}
\end{array}\right) ,\label{eq:9}
\end{equation}
 which satisfies the $\mu-\tau$ reflection symmetry. That means the first row elements are real, similar to $U_{HS}$.
 The $'+'$ and $'-'$  signs correspond to Case-I and Case-II respectively.\\
\indent~ To account the deviations from $\mu-\tau$ reflection symmetry we assume $\mu-\tau$ reflection symmetry to be preserved at the flavour symmetry scale $\Lambda_{\mu\tau}$ and consider its breaking due to RG running from $\Lambda_{\mu\tau}$ to electroweak scale $\Lambda_{EW}$. Below the high energy scale $\Lambda_{\mu\tau}$, the running of light effective Majorana neutrino mass matrix $M_\nu$ is described by the one loop RGE given by \cite{ chank,babu,casas,antusch,antusch1,chank1,ray}
\begin{equation}
16 \pi^2 \frac{dM_\nu}{dt}= \alpha M_\nu + C \left[ \lbrace(Y_l Y_l^{\dagger})M_\nu
+M_\nu (Y_l Y_l^{\dagger})^T \rbrace \right],\label{eq:c}
\end{equation}
where $t$ stands for $\ln(\mu/\mu_{\circ})$ with $\mu$ being the renormalization scale and $Y_l$ is the charged lepton Yukawa coupling matrix. The constants $C$ and $\alpha$ read as 
\begin{equation}
C=-\frac{3}{2},\ \ \alpha \simeq -3g_2^2 + 6 y_t^2 + \lambda \label{eq:d}
\end{equation}
for SM and 
\begin{equation}
 C= 1, \ \ \ \alpha \simeq -\frac{6}{5}g_1^2 -6 g_2^2 + 6 y_t^2 \label{eq:f}
\end{equation}
for MSSM, with $g_{1,2}$, $y_t$ and $\lambda$ being the gauge coupling constants, top quark Yukawa coupling constant and SM quartic Higgs coupling constant respectively. The integral solution to Eq.(\ref{eq:c}) is given by \cite{ingr1, ingr2,ohls,zhou,huang,zhang}
\begin{equation}
 M_{\nu}(\Lambda_{EW}) = I_{\alpha} I_l M_{\nu}(\Lambda_{\mu\tau}) I_l,\label{eq:11}
\end{equation}
where $M_{\nu}(\Lambda_{\mu\tau})$ represents the mass matrix that respects $\mu$-$\tau$ reflection symmetry (defined in Eq.(\ref{eq:10})), while $ M_{\nu}(\Lambda_{EW})$ represents the low energy mass matrix, which does not contain the symmetry. Further in Eq.(\ref{eq:11})
\begin{equation}
  \displaystyle I_\alpha = \exp \left[- \frac{C}{16 \pi^2} 
  \int_{\ln\left(\Lambda_{EW}/1GeV\right)}^{\ln\left(\Lambda_{\mu\tau}/1GeV\right)}\alpha(t)dt \right],\label{eq:12}
 \end{equation}
and $I_l=Diag\{I_e,I_{\mu},I_{\tau}\}$ with
\begin{equation}
   \displaystyle I_{e,\mu,\tau}  = \exp \left[- \frac{C}{16 \pi^2} 
              \int_{\ln\left(\Lambda_{EW}/1GeV\right)}^{\ln\left(\Lambda_{\mu\tau}/1GeV\right)} y_{e,\mu,\tau}^2(t)dt \right].\label{eq:13}
    \end{equation}
Since $y_e^2<<y_{\mu}^2<<y_{\tau}^2$, we may take the approximations: $I_e \approx 1$,
$I_{\mu}\approx 1$ and $I_{\tau}\approx 1+\epsilon$ such that
\begin{equation}
 \displaystyle \epsilon  = - \frac{C}{16 \pi^2} 
\int_{\ln\left(\Lambda_{EW}/1GeV\right)}^{\ln\left(\Lambda_{\mu\tau}/1GeV\right)} y_{\tau}^2(t)dt. \label{eq:14}
    \end{equation} 
With the above approximations, the low energy mass matrix at $\Lambda_{EW}$ in Eq.(\ref{eq:11}) becomes
\begin{equation}
M_{\nu} =I_{\alpha} \left(\begin{array}{ccc}
 a^{\mu\tau} & b^{\mu\tau} & (1+\epsilon)\left( b^{\mu\tau}\right)^{*}\\
 b^{\mu\tau} & c^{\mu\tau} & (1+\epsilon)d^{\mu\tau}\\
  (1+\epsilon)\left( b^{\mu\tau}\right)^{*} & (1+\epsilon)d^{\mu\tau} & (1+\epsilon)^2 \left( c^{\mu\tau}\right)^{*}\\
   \end{array}\right).\label{eq:15}
\end{equation}
In the leading order in $\epsilon$, $M_\nu$ can be expressed as 
\begin{equation}
    M_\nu \approx I_\alpha \left[ M_\nu^{\mu\tau}
    +\epsilon \Delta M_\nu\right],\label{eq:K}
\end{equation}
where the reflection symmetric mass matrix $M_\nu^{\mu\tau}$ is given in Eq.(\ref{eq:10}) and $\Delta M_\nu $ is given by
\begin{equation}
{\Delta M}_\nu = \left(\begin{array}{ccc}
 0 & 0 & \left( b^{\mu\tau}\right)^{*}\\
 0 & 0 & d^{\mu\tau}\\
 \left( b^{\mu\tau}\right)^{*} & d^{\mu\tau} & 2\left( c^{\mu\tau}\right)^{*} \\
 \end{array}\right).\label{eq:L}
\end{equation}
 $M_\nu$ can be diagonalized by the lepton mixing matrix in Eq.(\ref{eq:b}) as
\begin{equation}
  U^{\dagger} M_\nu U^* = D_{\nu}= Diag(m_1, m_2, m_3).\label{eq:16}
\end{equation}
Similarly,  $U^{\mu\tau}$ diagonalizes the symmetric mass matrix as
\begin{equation}
\left( U^{\mu\tau}\right)^\dagger M_\nu^{\mu\tau}\left( U^{\mu\tau}\right)^*=D_\nu^{\mu\tau} = Diag({m_1^{\mu\tau}, m_2^{\mu\tau}, m_3^{\mu\tau})}.\label{eq:17}
\end{equation}
Now using Eqs.(\ref{eq:16}) and (\ref{eq:17}) in  Eq.(\ref{eq:K}) and following algebraic calculations, we can obtain the analytical relations between the mass eigenvalues, mixing angles and CP phases corresponding to the energy scales $\Lambda_{\mu\tau}$ and $\Lambda_{EW}$ \cite{pegu}. The corresponding expressions for mass eigenvalues are 
\begin{equation}
m_1 \approx I_{\alpha} \left[ 1 + \epsilon \left[ \left( c_{12}^{\mu\tau}\right)^2 \left( s_{13}^{\mu\tau}\right)^2 +\left( s_{12}^{\mu\tau}\right)^2 \right] \right] m_{1}^{\mu\tau},\label{eq:18}
\end{equation}
\begin{equation}
 m_2 \approx I_{\alpha} \left[ 1 + \epsilon \left[ \left( s_{12}^{\mu\tau}\right)^2 \left( s_{13}^{\mu\tau}\right)^2 + \left( c_{12}^{\mu\tau}\right)^2 \right] \right] m_{2}^{\mu\tau},\label{eq:19}
\end{equation}
\begin{equation}
m_3 \approx I_{\alpha} \left[ 1 + \epsilon \left( c_{13}^{\mu\tau}\right)^2  \right] m_{3}^{\mu\tau},\label{eq:20}
\end{equation} 
 \cite{pegu}. The R.H.S of the Eqs.(\ref{eq:18}-\ref{eq:20}) represent the high energy mass eigenvalues and mixing angles, while L.H.S represent the low energy mass eigenvalues. Similarly, corresponding low energy mixing angles in terms of high energy mass eigenvalues and mixing angles are
\begin{equation}
\displaystyle \sin^2\theta_{13} \approx \left(s_{13}^{\mu\tau}\right)^2 - \epsilon \left(s_{13}^{\mu\tau}\right)^2 c_{13}^{\mu\tau} \left[\left(c_{12}^{\mu\tau}\right)^2 \Delta_{31} + \left(s_{12}^{\mu\tau}\right)^2 \Delta_{32}\right],\label{eq:21}
 \end{equation}
\begin{equation}
\displaystyle \sin^2\theta_{23} \approx \frac{1}{2} -\frac{\epsilon}{2}\left[ \left(s_{12}^{\mu\tau}\right)^2 \Delta_{31}^{-1} +\left(c_{12}^{\mu\tau}\right)^2 \Delta_{32}^{-1}\right],\label{eq:22}
\end{equation}
\begin{equation}
\displaystyle \sin^2\theta_{12} \approx \left(s_{12}^{\mu\tau}\right)^2 - \epsilon \left(s_{12}^{\mu\tau}\right)^2 c_{12}^{\mu\tau} \left[\left(s_{13}^{\mu\tau}\right)^2 \left( \Delta_{31}-\Delta_{32}\right) + \left(c_{13}^{\mu\tau}\right)^2 \Delta_{21}\right]
,\label{eq:23}
\end{equation}
 \cite{pegu}, where $\displaystyle \Delta_{21}= \frac{m_2^{\mu\tau} + m_1^{\mu\tau}}{m_2^{\mu\tau} - m_1^{\mu\tau}}, \ \
\Delta_{31}= \frac{m_3^{\mu\tau} + m_1^{\mu\tau}}{m_3^{\mu\tau} - m_1^{\mu\tau}}, \ \
 \Delta_{32}= \frac{m_3^{\mu\tau} + m_2^{\mu\tau}}{m_3^{\mu\tau} - m_2^{\mu\tau}}$. Further, the expressions for low energy CP phases are
\begin{equation}
\delta^I/\delta^{II} = (\pi/2)/(3\pi/2) + T_{e1}+ T_{e2} - T_{e3} + T_{\mu 3} + T_{\tau 3}, \label{eq:24}
\end{equation}
\begin{equation}
\alpha^I/\alpha^{II} = (\pi/2)/(3\pi/2) - T_{e2} - T_{\mu3} - T_{\tau3}, \label{eq:25}
\end{equation}
\begin{equation}
\beta^I/\beta^{II} = (\pi/2)/(3\pi/2) - T_{e1} - T_{\mu3} - T_{\tau3}, \label{eq:26}
\end{equation}
where 
\begin{equation}
\begin{aligned}
T_{e1}= \arctan \left[\frac{\pm \frac{\displaystyle\epsilon}{2}\left[\Delta_{31}^{-1}-
 \Delta_{21}^{-1}\right] s_{12}^{\mu\tau}s_{13}^{\mu\tau}}{c_{12}^{\mu\tau}+ \frac{\displaystyle\epsilon}{2} \left[ \left(s_{13}^{\mu\tau}\right)^2 \Delta_{31}+\left(s_{12}^{\mu\tau}\right)^2 \left(c_{13}^{\mu\tau}\right)^2\Delta_{21}
 \right] c_{12}^{\mu\tau}} \right],
 \end{aligned}\label{eq:27}
\end{equation}
\begin{equation}
\begin{aligned}
T_{e2}= \arctan \left[\frac{\mp \frac{\displaystyle\epsilon}{2}\left[\Delta_{32}^{-1}+
 \Delta_{21}^{-1}\right] c_{12}^{\mu\tau}s_{13}^{\mu\tau}}{s_{12}^{\mu\tau}+ \frac{\displaystyle \epsilon}{2} \left[ \left(s_{13}^{\mu\tau}\right)^2 \Delta_{32}-
\left(c_{12}^{\mu\tau}\right)^2 \left(c_{13}^{\mu\tau}\right)^2 \Delta_{21} \right] s_{12}^{\mu\tau}} \right],
\end{aligned}\label{eq:28}
\end{equation}
\begin{equation}
\begin{aligned}
T_{e3}= \arctan \left[\frac{\mp \frac{\displaystyle\epsilon}{2}\left[\Delta_{32}^{-1}-
 \Delta_{31}^{-1}\right]s_{12}^{\mu\tau} c_{12}^{\mu\tau}\left(c_{13}^{\mu\tau}\right)^2 }{s_{13}^{\mu\tau}- \frac{\displaystyle\epsilon}{2} \left[ \left(s_{12}^{\mu\tau}\right)^2 \Delta_{32}+
\left(c_{12}^{\mu\tau}\right)^2 \Delta_{31} \right] s_{13}^{\mu\tau}\left(c_{13}^{\mu\tau}\right)^2}  \right],
\end{aligned}\label{eq:29}
\end{equation}
\begin{equation}
\begin{aligned}
T_{\mu3}= \arctan \left[\frac{\pm \frac{\displaystyle\epsilon}{2}\left[\Delta_{32}-
 \Delta_{31}+\Delta_{32}^{-1}-\Delta_{31}^{-1}\right]s_{12}^{\mu\tau} c_{12}^{\mu\tau}s_{13}^{\mu\tau}}{1+ \frac{\displaystyle\epsilon}{2} \left[ \left(s_{12}^{\mu\tau}\right)^2 \Delta_{32}+
\left(c_{12}^{\mu\tau}\right)^2 \Delta_{31} \right] \left(s_{13}^{\mu\tau}\right)^2-\left[ \left(c_{12}^{\mu\tau}\right)^2 \Delta_{32}^{-1}+
\left(s_{12}^{\mu\tau}\right)^2 \Delta_{31}^{-1} \right]}  \right],
\end{aligned}\label{eq:30}
\end{equation}
\begin{equation}
\begin{aligned}
T_{\tau3}= \arctan \left[\frac{\mp \frac{\displaystyle\epsilon}{2}\left[\Delta_{32}-
 \Delta_{31}-\Delta_{32}^{-1}+\Delta_{31}^{-1}\right]s_{12}^{\mu\tau} c_{12}^{\mu\tau}s_{13}^{\mu\tau}}{1+ \frac{\displaystyle\epsilon}{2} \left[ \left(s_{12}^{\mu\tau}\right)^2 \Delta_{32}+
\left(c_{12}^{\mu\tau}\right)^2 \Delta_{31} \right] \left(s_{13}^{\mu\tau}\right)^2+\left[ \left(c_{12}^{\mu\tau}\right)^2 \Delta_{32}^{-1}+
\left(s_{12}^{\mu\tau}\right)^2 \Delta_{31}^{-1} \right]}  \right].
\end{aligned}\label{eq:31}
\end{equation} 
The superscripts $'I'$ and $'II'$ in the CP phases correspond to Case-I $(\theta_{23}^{\mu\tau}=\pi/4,\  \delta^{\mu\tau}= \pi/2)$ and Case-II$(\theta_{23}^{\mu\tau}=\pi/4,\  \delta^{\mu\tau}= 3\pi/2)$ respectively. Also in Eqs.(\ref{eq:27}-\ref{eq:31}) $'+'$ and $'-'$ signs correspond to Case-I  and Case-II respectively.
\section{Numerical analysis and results}
\indent~ In this section, we numerically estimate the values of $|m_{ee}|$ at the electroweak scale, which is chosen at the top quark mass scale: $\Lambda_{EW}=m_t=172.76\ GeV$ using  Eq. (\ref{eq:5}). From this equation, it is clear that the value of
$|m_{ee}|$ depends on eight low energy parameters, namely three mass eigenvalues ($m_1,\ m_2,\ m_3$), two mixing angles ($\theta_{12},\ \theta_{13}$) and three CP violating phases ($\delta,\ \alpha,\ \beta$). Therefore, to compute $|m_{ee}|$ at $\Lambda_{EW}$, we first determine the corresponding low-energy values of all these parameters. We obtain these low energy parameters using Eqs. (\ref{eq:18}-\ref{eq:26}). In these equations all the parameters appearing in the right-hand side with superscript $'\mu\tau'$ represent the high energy values at the flavor symmetry scale $\Lambda_{\mu\tau}$, while the left-hand side represents corresponding low energy values at $\Lambda_{EW}$. These equations also involve the parameters $I_\alpha$ and $\epsilon$. We compute the values of these two parameters using a numerical approximation method. To carry out the numerical analysis the flavor symmetry scale is set at the seesaw scale: $\Lambda_{\mu\tau}=10^{14}\ GeV$. The values of all the parameters ($m_1^{\mu\tau},\ m_2^{\mu\tau},\ m_3^{\mu\tau},\ \theta_{12}^{\mu\tau},\ \theta_{23}^{\mu\tau},\ \theta_{13}^{\mu\tau},\ \delta^{\mu\tau},\ \alpha^{\mu\tau},\ \beta^{\mu\tau}$) 
at $\Lambda_{\mu\tau}$ are taken as inputs and corresponding low energy predictions due to radiative corrections are computed using Eqs. (\ref{eq:18}-\ref{eq:26}). As we work in the Minimal Supersymmetric framework, we consider three different SUSY breaking scales: $\Lambda_s= 1 \ TeV$, $7 \ TeV$ and $14 \ TeV$ and also two different values of $\tan\beta$, namely $30$ and $58$ to perform the analysis. Since there exist two possible scenarios regarding the order of mass eigenvalues- Normal order (NO) ($m_1<m_2<m_3$) and Inverted order (IO) ($m_3<m_1<m_2$), we perform the analysis separately for each scenario. 
Again, in each scenario of NO and IO, the numerical results are presented separately for the two different choices of $\tan\beta=30$ and $58$.\\
\indent~ We first present the calculated values of $I_\alpha$ and $\epsilon$ for $\tan\beta=30$ and $58$ with $\Lambda_s= 1 \ TeV$, $7 \ TeV$ and $14 \ TeV$ in Table 1, which have been taken from our earlier work \cite{pegu}. Using these values of $I_\alpha$ and $\epsilon$ and choosing high energy inputs of all the neutrino parameters in Eqs. (\ref{eq:18}-\ref{eq:26}), we estimate the corresponding low-energy mass eigenvalues and mixing parameters. Since the $\mu$-$\tau$ reflection symmetry is assumed to be preserved at $\Lambda_{\mu\tau}$, the input values of $\theta_{23}^{\mu\tau}$, $\delta^{\mu\tau},\ \alpha^{\mu\tau}$ and $\beta^{\mu\tau}$  are constrained as given in Eqs.(\ref{eq:7}-\ref{eq:8}) due to reflection symmetry while other parameters ($m_1^{\mu\tau},\ m_2^{\mu\tau},\ m_3^{\mu\tau},\ \theta_{12}^{\mu\tau},\ \theta_{13}^{\mu\tau}$) remain as free. The input values of free parameters are chosen such that the low-energy predictions are consistent with neutrino oscillation data (mass squared differences, mixing angles and Dirac CP phase) and the sum of mass eigenvalues $\Sigma \vert{m_i}\vert < 0.12\ eV$ ($i=1, 2, 3$) provided by cosmological observations. In NO scenario, the chosen high energy inputs for $\tan\beta=30$ and $58$ are taken from our earlier work \cite{pegu}. Similarly, in the IO scenario, the chosen high energy inputs of the free parameters for $\tan\beta=58$ are also taken from \cite{pegu}. However for $\tan\beta=30$, the choice of the high energy input values of the free parameters are determined in the present study, which was not covered in \cite{pegu}. Finally, using low energy values obtained as a result of radiative breaking of $\mu$-$\tau$ reflection symmetry,  we compute the effective Majorana neutrino mass $|m_{ee}|$ at the electroweak scale $\Lambda_{EW}$.
\begin{table}[H]
\centering
\begin{tabular}{lccc|ccc}
\hline
 & \multicolumn{3}{c|}{$\tan\beta=30$} & \multicolumn{3}{c}{$\tan\beta=58$} \\
\hline
Parameter & $1\ \mathrm{TeV}$ & $7\ \mathrm{TeV}$ & $14\ \mathrm{TeV}$ & $1\ \mathrm{TeV}$ & $7\ \mathrm{TeV}$ & $14\ \mathrm{TeV}$ \\
\hline
$I_\alpha$ & 0.884058 & 0.93213 & 0.945856 & 0.829346 & 0.899076 & 0.921216\\

$\epsilon$ & -0.012505 &-0.012035&-0.011823 & -0.073143 & -0.064597 & -0.062217\\
\hline
\end{tabular}\caption{ Calculated values of $I_\alpha$ and $\epsilon$ for $\tan\beta= 30$ and $58$ with $\Lambda_s=  1\ TeV$, $7\ TeV$, $14\ TeV$.}
\end{table}
\subsection{Normal order ($m_1<m_2<m_3$)}
\indent~ In NO scenario, we choose the high energy input values of the mass eigenvalues such that $m_1<m_2<m_3$. We first present the numerical results for the case of $\tan\beta= 30$. The high energy input values and corresponding low energy output values of the parameters for all three SUSY breaking scales  $\Lambda_s=  1\ TeV$, $7\ TeV$ and $14\ TeV$  are presented in Table 2. The table consists of three vertical blocks, where the first, second and third blocks respectively contain the results corresponding to $\Lambda_s=  1\ TeV$, $7\ TeV$ and $14\ TeV$. In each block, the left column represents the high energy input values and the right column represents the low energy output values. Since $\mu$–$\tau$ reflection symmetry is assumed to be preserved at $\Lambda_{\mu\tau}$, therefore among the given high-energy parameters, the values of $\theta_{23}^{\mu\tau}$ and CP phases are constrained by this symmetry. Further, it allows two possible values of the Dirac CP phase. In light of this, we consider two cases: Case-I ($\delta^{\mu\tau}= \pi/2,\ \alpha^{\mu\tau}=\beta^{\mu\tau}=3 \pi/2$) and Case-II ($\delta^{\mu\tau}= 3\pi/2,\ \alpha^{\mu\tau}= \beta^{\mu\tau}= \pi/2$) as discussed in section 2. In the horizontal direction, Table 2 is also divided into three blocks, where the first block contains values of mass eigenvalues and mixing angles. Again, the second and third blocks contain the values of CP phases corresponding to Case-I and Case-II. Finally, using the low energy predictions of mass eigenvalues and mixing parameters in Eq. (\ref{eq:5}), we estimate the value of $|m_{ee}|$. The calculated values of $|m_{ee}|$ are shown in the second and third blocks of Table 2 in the horizontal direction. For each choice of SUSY breaking scale, the values of $|m_{ee}|$ are found to be equal in both Case-I and Case-II. For $\Lambda_s=  1\ TeV$, $7\ TeV$ and $14\ TeV$ we get $|m_{ee}|= 0.024189\ eV$, \ $0.025361\ eV$ and $0.026337\ eV$ respectively. Note that the most stringent upper bound on $|\langle m \rangle_{ee}|$ is currently provided by the KamLAND-Zen Collaboration, which reports  $|\langle m \rangle_{ee}|<(0.028-0.122)\ eV$. Thus the predictions on $|m_{ee}|$ for each choice of $\Lambda_s$ with $\tan\beta=30$ are found to be consistent with the current experimental upper bound. Specifically, we note that all these predictions lie below the interval $(0.028 - 0.122)\ eV$. Further we observe that the prediction on $|m_{ee}|$ increases as the SUSY breaking scale $\Lambda_s$ increases.\\
\indent~ The numerical results for $\tan\beta = 58$ with $\Lambda_s = 1\ TeV$, $7\ TeV$ and $14\ TeV$ are presented in Table 3 in a similar manner as like Table 2. The high energy input values and corresponding low energy values of all the parameters for each choice of $\Lambda_s$ are shown in three vertical blocks. The calculated values of $|m_{ee}|$ at $\Lambda_{EW}$ are presented in the second and third blocks in the horizontal direction respectively for Case-I and Case-II. As observed in the previous case ($\tan\beta = 30$), the predictions on $|m_{ee}|$ in this case are also found to be equal for both Case-I and Case-II. We obtain $|m_{ee}|=0.013733\ eV$, $0.030209$ and $0.031013\ eV$ for $\Lambda_s = 1\ TeV$, $7\ TeV$ and $14\ TeV$ respectively. All these predictions are consistent with the experimental upper bound. Specifically, we note that the prediction on $|m_{ee}|$ for $\Lambda_s = 1\ TeV$ ($0.013733\ eV$) is lying below the interval $(0.028 - 0.122)$ while for $\Lambda_s = 7\ TeV$ and $14\ TeV$ the predictions ($0.030209\ eV,\ 0.031013\ eV$) are lying within the interval. Regarding the effects of the variation of SUSY breaking scale, we see that, for $\tan\beta = 58$ the predictions on  $|m_{ee}|$ increase as the $\Lambda_s$ increases. This observation is similar to that found in the case of $\tan\beta = 30$.
\begin{table}[t]
\centering
\begin{tabular}{l|cc|cc|cc}
\hline
& \multicolumn{2}{c|}{$\Lambda_s = 1\ \mathrm{TeV}$} & \multicolumn{2}{c|}{$\Lambda_s = 7\ \mathrm{TeV}$} & \multicolumn{2}{c}{$\Lambda_s = 14\ \mathrm{TeV}$} \\
\hline
Parameter & Input & Output & Input & Output & Input & Output \\
\hline
$m_1\ (\mathrm{eV})$ & 0.024101 & 0.022938 & 0.022456 & 0.024147 & 0.022546 & 0.025145 \\
$m_2\ (\mathrm{eV})$ & 0.025961 & 0.024583 & 0.024015 & 0.025691 & 0.023992 & 0.026629 \\
$m_3\ (\mathrm{eV})$ & 0.058445 & 0.055139 & 0.052240 & 0.055674 & 0.051000 & 0.056372 \\
$\sin^2\theta_{12}$ & 0.3025 & 0.3382 & 0.3025 & 0.3448 & 0.3025 & 0.3484 \\
$\sin^2\theta_{23}$ & 0.5 & 0.5001 & 0.5 & 0.5003 & 0.5 & 0.5004 \\
$\sin^2\theta_{13}$ & 0.02161 & 0.02194 & 0.02161 & 0.02198 & 0.02161 & 0.02199 \\
\hline
\multicolumn{7}{l}{\textbf{Case-I}}\\
\hline
$\delta\ (^\circ)$ & 90 & 90 & 90 & 90 & 90 & 90 \\
$\alpha\ (^\circ)$ & 270 & 270 & 270 & 270 & 270 & 270 \\
$\beta\ (^\circ)$ & 270 & 270 & 270 & 270 & 270 & 270 \\
\hline
$|m_{ee}|\ (\mathrm{eV})$ & -- & 0.024189 & -- & 0.025361 & -- & 0.026337 \\
\hline
\multicolumn{7}{l}{\textbf{Case-II}}\\
\hline
$\delta\ (^\circ)$ & 270 & 270 & 270 & 270 & 270 & 270 \\
$\alpha\ (^\circ)$ & 90 & 90 & 90 & 90 & 90 & 90 \\
$\beta\ (^\circ)$ & 90 & 90 & 90 & 90 & 90 & 90 \\
\hline
$|m_{ee}|\ (\mathrm{eV})$ & -- & 0.024189 & -- & 0.025361 & -- & 0.026337 \\
\hline
\end{tabular}
\caption{High energy input values of parameters at $\Lambda_{\mu\tau} = 10^{14}\ GeV$ and corresponding low energy values and the computed values of $|m_{ee}|$ at $\Lambda_{EW} = 172.76\ GeV$ in NO scenario for $\tan\beta = 30$. In the vertical direction, first, second and third blocks respectively represent the values corresponding to $\Lambda_s = 1\ TeV$, $7\ TeV$ and $14\ TeV$. In the horizontal direction, second and third blocks represent the values of CP phases and $|m_{ee}|$ corresponding to Case-I and Case-II respectively.
}
\end{table}

\begin{table}[H]
\centering
\begin{tabular}{l|cc|cc|cc}
\hline
& \multicolumn{2}{c|}{$\Lambda_s = 1\ \mathrm{TeV}$} & \multicolumn{2}{c|}{$\Lambda_s = 7\ \mathrm{TeV}$} & \multicolumn{2}{c}{$\Lambda_s = 14\ \mathrm{TeV}$} \\
\hline
Parameter & Input & Output & Input & Output & Input & Output \\
\hline
$m_1\left(eV\right)$ & 0.014523& 0.012708 & 0.028796 & 0.029283 & 0.029548& 0.029392 \\
$m_2\left(eV\right)$ &0.018009& 0.015287& 0.031420& 0.030081 &0.031420 &0.032171\\
$m_3\left(eV\right)$ & 0.062765& 0.052109 & 0.060100&0.058238 & 0.058066   & 0.059110 \\
$\sin^2\theta_{12}$ & 0.3025& 0.3303 & 0.3025 & 0.3081& 0.3025 & 0.3436 \\
$\sin^2\theta_{23}$ & 0.5& 0.5009& 0.5& 0.5008& 0.5   & 0.5009 \\
$\sin^2\theta_{13}$ & 0.02161& 0.02295 & 0.02161& 0.02371 & 0.02161  & 0.02316\\
\hline
\multicolumn{7}{l}{\textbf{Case-I}}\\
\hline
$\delta\left(/^\circ\right)$ & 90 & 90& 90& 90 &90 &90\\
$\alpha\left(/^\circ\right)$ & 270& 270&270 & 270 &270& 270\\
$\beta\left(/^\circ\right)$ & 270 &270& 270 & 270& 270& 270\\
\hline
$|m_{ee}|\left(eV\right)$  & -& 0.013733 &-& 0.030209 &- & 0.031013\\
\hline
\multicolumn{7}{l}{\textbf{Case-II}}\\
\hline
$\delta\left(/^\circ\right)$ & 270& 270& 270 & 270 &270&270  \\
$\alpha\left(/^\circ\right)$ & 90& 90& 90 & 90& 90&90  \\
$\beta\left(/^\circ\right)$ & 90& 90 & 90 & 90 &90& 90\\
\hline
$|m_{ee}|\left(eV\right)$  &-& 0.013733& - & 0.030209 &- & 0.031013\\
\hline
\end{tabular}
\caption{High energy input values of parameters at $\Lambda_{\mu\tau} = 10^{14}\ GeV$ and corresponding low energy values and the computed values of $|m_{ee}|$ at $\Lambda_{EW} = 172.76\ GeV$ in NO scenario for $\tan\beta = 58$. In the vertical direction, first, second and third blocks respectively represent the values corresponding to $\Lambda_s = 1\ TeV$, $7\ TeV$ and $14\ TeV$. In the horizontal direction, second and third blocks represent the values of CP phases and $|m_{ee}|$ corresponding to Case-I and Case-II respectively.
}
\end{table}
\subsection{ Inverted order ($m_3<m_1<m_2$)}
\indent~ We perform the numerical analysis for the IO scenario in the same manner as done for the NO scenario. Here, the high-energy mass eigenvalues are taken in the order $m_3<m_1<m_2$. We first present the numerical results for $\tan\beta=30$ followed by those for $\tan\beta=58$. The numerical results for both cases with $\Lambda_s = 1\ TeV$, $7\ TeV$ and $14\ TeV$ are presented in Tables 4 and 5 respectively. The presentation structure of this table is same as those of Tables 2 and 3. The high energy input values and corresponding low energy values of all the parameters for $\tan\beta=30$ with $\Lambda_s = 1\ TeV$, $7\ TeV$ and $14\ TeV$ are presented in three vertical blocks of Table 4. The calculated values of $|m_{ee}|$ are shown in the second and third blocks in the horizontal direction respectively for Case-I and Case-II. As found in the NO scenario, the predictions for $|m_{ee}|$ are same in both Case-I and II for each value of $\Lambda_s$. Numerically we obtain $|m_{ee}|= 0.055238\ eV$, $0.049962\ eV$ and $0.049851\ eV$ for three SUSY breaking scales respectively. We observe that these predictions are consistent with the experimental upper bound $|m_{ee}|<(0.028-0.122)\ eV$. We note that these predictions of $|m_{ee}|$ are larger as compared to those obtained in the NO scenario. Significantly, we observe that the predictions of $|m_{ee}|$ in the NO scenario for $\tan\beta=30$ are lying below the interval $(0.028-0.122)\ eV$, however in the present IO scenario, the predictions are lying within the interval. Regarding the effects of variation of SUSY breaking scales we see that in the present IO scenario for $\tan\beta=30$, the predictions of $|m_{ee}|$ decrease as $\Lambda_s$ increases. Significantly, this nature of effects of variation of $\Lambda_s$ is opposite to that observed in the NO scenario, where value of $|m_{ee}|$ increases with the increase of $\Lambda_s$.
\begin{table}[H]
\centering
\begin{tabular}{l|cc|cc|cc}
\hline
& \multicolumn{2}{c|}{$\Lambda_s = 1\ \mathrm{TeV}$} & \multicolumn{2}{c|}{$\Lambda_s = 7\ \mathrm{TeV}$} & \multicolumn{2}{c}{$\Lambda_s = 14\ \mathrm{TeV}$} \\
\hline
Parameter & Input & Output & Input & Output & Input & Output \\
\hline
$m_1\left(eV\right)$ & 0.053299 & 0.050804 & 0.046399& 0.050677 &  0.045287 & 0.050586\\
$m_2\left(eV\right)$ &0.054510 & 0.051567 & 0.048199& 0.051380 &0.046270 &0.051264\\
$m_3\left(eV\right)$ & 0.008899 & 0.008396 & 0.007001& 0.007461 & 0.006001& 0.006633 \\
$\sin^2\theta_{12}$ & 0.3025& 0.2765 &0.3176& 0.2867& 0.3025 & 0.2904 \\
$\sin^2\theta_{23}$ & 0.5& 0.4997& 0.5& 0.4994& 0.5& 0.4993 \\
$\sin^2\theta_{13}$ & 0.02161& 0.02112 & 0.02161& 0.02110 & 0.02161& 0.02111\\
\hline
\multicolumn{7}{l}{\textbf{Case-I}}\\
\hline
$\delta\left(/^\circ\right)$ & 90 & 90& 90& 90 &90 &90\\
$\alpha\left(/^\circ\right)$ & 270& 270&270 & 270 &270& 270\\
$\beta\left(/^\circ\right)$ & 270 &270& 270 & 270& 270& 270\\
\hline
$|m_{ee}|\left(eV\right)$  &-& 0.055238 &-& 0.049962 &- & 0.049851\\
\hline
\multicolumn{7}{l}{\textbf{Case-II}}\\
\hline
$\delta\left(/^\circ\right)$ & 270& 270& 270 & 270 &270&270  \\
$\alpha\left(/^\circ\right)$ & 90& 90& 90 & 90& 90&90  \\
$\beta\left(/^\circ\right)$ & 90& 90 & 90 & 90 &90& 90\\
\hline
$|m_{ee}|\left(eV\right)$  &-&0.055238&-& 0.049962&- & 0.049851\\
\hline
\end{tabular}
\caption{High energy input values of parameters at $\Lambda_{\mu\tau} = 10^{14}\ GeV$ and corresponding low energy values and the computed values of $|m_{ee}|$ at $\Lambda_{EW} = 172.76\ GeV$ in IO scenario for $\tan\beta = 30$. In the vertical direction, first, second and third blocks respectively represent the values corresponding to $\Lambda_s = 1\ TeV$, $7\ TeV$ and $14\ TeV$. In the horizontal direction, second and third blocks represent the values of CP phases and $|m_{ee}|$ corresponding to Case-I and Case-II respectively.
}\end{table}
 Turning to the case of $\tan\beta=58$, the high energy input values and corresponding low energy values of parameters for $\Lambda_s = 1\ TeV$, $7\ TeV$ and $14\ TeV$ are presented in the first, second and third vertical blocks of Table 5 respectively. The calculated  low energy values of $|m_{ee}|$ for Case-I and Case-II are presented in the second and third horizontal blocks respectively. Again, as observed in the previous analysis, for both Case-I and Case-II, the predicted values of $|m_{ee}|$ for each choice of $\Lambda_s$ are found to be equal. We obtain that the values of $|m_{ee}|$ are $0.051176\ eV$, $0.050662\ eV$ and $0.050084\ eV$ for the three choices of $\Lambda_s$ respectively. The consistency of these predictions with the current experimental upper bound on the effective Majorana neutrino mass remains similar to that observed in the previous analysis. We note that for $\tan\beta=58$, all predicted values  of $|m_{ee}|$ are larger than those obtained in the NO scenario as well as those in the present IO scenario for $\tan\beta=30$. Further, the nature of effects of variation of SUSY breaking scale is similar to that observed in present IO scenario for $\tan\beta=30$, where $|m_{ee}|$ decreases with increase of $\Lambda_s$.
\begin{table}[t]
\centering
\begin{tabular}{l|cc|cc|cc}
\hline
& \multicolumn{2}{c|}{$\Lambda_s = 1\ \mathrm{TeV}$} & \multicolumn{2}{c|}{$\Lambda_s = 7\ \mathrm{TeV}$} & \multicolumn{2}{c}{$\Lambda_s = 14\ \mathrm{TeV}$} \\
\hline
Parameter & Input & Output & Input & Output & Input & Output \\
\hline
$m_1\left(eV\right)$ & 0.057862& 0.051714 & 0.049885& 0.050735 &  0.047981 & 0.051215\\
$m_2\left(eV\right)$ &0.061499 & 0.052421& 0.052870& 0.051464 &0.050775 &0.051932\\
$m_3\left(eV\right)$ & 0.017001& 0.014328 & 0.010001& 0.009691 & 0.012001 & 0.012217 \\
$\sin^2\theta_{12}$ & 0.3025& 0.3332 &0.3176& 0.3025& 0.3025 & 0.3253 \\
$\sin^2\theta_{23}$ & 0.5& 0.4991& 0.5& 0.4992& 0.5& 0.4979 \\
$\sin^2\theta_{13}$ & 0.02161& 0.02055 & 0.02161& 0.02138 & 0.02161& 0.02118\\
\hline
\multicolumn{7}{l}{\textbf{Case-I}}\\
\hline
$\delta\left(/^\circ\right)$ & 90 & 90& 90& 90 &90 &90\\
$\alpha\left(/^\circ\right)$ & 270& 270&270 & 270 &270& 270\\
$\beta\left(/^\circ\right)$ & 270 &270& 270 & 270& 270& 270\\
\hline
$|m_{ee}|\left(eV\right)$  &-& 0.051176 &-& 0.050662 &- & 0.050084\\
\hline
\multicolumn{7}{l}{\textbf{Case-II}}\\
\hline
$\delta\left(/^\circ\right)$ & 270& 270& 270 & 270 &270&270  \\
$\alpha\left(/^\circ\right)$ & 90& 90& 90 & 90& 90&90  \\
$\beta\left(/^\circ\right)$ & 90& 90 & 90 & 90 &90& 90\\
\hline
$|m_{ee}|\left(eV\right)$  &-&0.051176&-& 0.050662&- & 0.050084\\
\hline
\end{tabular}
\caption{High energy input values of parameters at $\Lambda_{\mu\tau} = 10^{14}\ GeV$ and corresponding low energy values and the computed values of $|m_{ee}|$ at $\Lambda_{EW} = 172.76\ GeV$ in IO scenario for $\tan\beta = 58$. In the vertical direction, first, second and third blocks respectively represent the values corresponding to $\Lambda_s = 1\ TeV$, $7\ TeV$ and $14\ TeV$. In the horizontal direction, second and third blocks represent the values of CP phases and $|m_{ee}|$ corresponding to Case-I and Case-II respectively.
}
\end{table}
\section{Summary and discussion}
\indent~ Neutrinoless double beta ($0\nu\beta\beta$) decay is the only process that can directly tell us that neutrinos are Majorana in nature. The rate of this decay depends on the effective Majorana neutrino mass $m_{ee}$. The KamLAND-Zen collaboration provides an upper bound on $|m_{ee}|$. Their latest result gives $|m_{ee}|<(0.028-0.122)\ eV$. In this work, we study the prediction of $|m_{ee}|$ at the electroweak scale $\Lambda_{EW}$ under the scenario of deviation from $\mu$–$\tau$ reflection symmetry due to radiative corrections. We employ the general expression for $|m_{ee}|$ given in Eq.(\ref{eq:5}), which involves three mass eigenvalues, the mixing angles $\theta_{12}$,\ $\theta_{13}$ and CP phases $\delta$,\ $\alpha$ and $\beta$. The low energy values of these parameters are determined as a corrections to $\mu$–$\tau$ reflection symmetry due to RG running. The high energy values of the mass eigenvalues and $\theta_{12}$ and $\theta_{13}$ are chosen as free parameters, while those of CP phases are constrained by $\mu$–$\tau$ reflection symmetry, which is assumed to be preserved  at the high energy scale. Using the low energy values of parameters determined in this way, we predict the values of $|m_{ee}|$ using Eq.(\ref{eq:5}). Working in the MSSM framework, we choose three different SUSY breaking scales at $\Lambda_s= 1 \ TeV$, $7 \ TeV$ and $14 \ TeV$ and two different values of $\tan\beta= 30$ and $58$ to perform the analysis. Further, we carry out the numerical analysis for both the scenarios of NO and IO. In the entire analysis, the predictions of $|m_{ee}|$ are found to be equal in both Case-I ($\delta^{\mu\tau}= \pi/2,\ \alpha^{\mu\tau}=\beta^{\mu\tau}=3 \pi/2$) and Case-II ($\delta^{\mu\tau}= 3\pi/2,\ \alpha^{\mu\tau}= \beta^{\mu\tau}= \pi/2$). We obtain $|m_{ee}|= 0.024189\ eV$, $0.025361\ eV$ and $0.026337\ eV$ respectively for $\Lambda_s= 1 \ TeV$, $7 \ TeV$ and $14 \ TeV$ in the NO scenario with $\tan\beta= 30$. For $\tan\beta= 58$ in the same NO scenario the calculated values of $|m_{ee}|$ are $0.013733\ eV$, $0.030209\ eV$ and $0.031013\ eV$ corresponding to $\Lambda_s= 1 \ TeV$, $7 \ TeV$ and $14 \ TeV$ respectively. In the IO scenario, we obtain values of $|m_{ee}|=0.055238\ eV$, $0.049962\ eV$  and $0.049851\ eV$ for $\tan\beta= 30$ and for $\tan\beta= 58$, we get $|m_{ee}|=0.051176\ eV$, $0.050662\ eV$ and $0.050084\ eV$ corresponding to $\Lambda_s= 1 \ TeV$, $7 \ TeV$ and $14 \ TeV$ respectively. All these predictions are consistent with the current upper bound $|m_{ee}| < (0.028 - 0.122)\ eV$ set by the KamLAND-Zen collaboration. We note that in the NO scenario, all the predictions for $\tan\beta = 30$ lie below the interval $(0.028 - 0.122)\ eV$, whereas for $\tan\beta = 58$, all the predictions lie within the interval, except the one corresponding to $\Lambda_s = 1\ TeV$. On the other hand, in the IO scenario, all predicted values of $|m_{ee}|$ lie within this interval for both $\tan\beta = 30$ and $\tan\beta = 58$. We also observe that the predicted values of $|m_{ee}|$ in the IO scenario are larger as compared to those observed in the NO scenario. Since we have considered three choices of SUSY breaking scales $\Lambda_s$ , it allows us to study the effects of variation of $\Lambda_s$ on the prediction of $|m_{ee}|$. We observe that in NO scenario, values of $|m_{ee}|$ increases as $\Lambda_s$ increases, while the nature of effects of variation of $\Lambda_s$ is found to be opposite in the IO scenario. In the IO scenario, the predictions of $|m_{ee}|$ decreases with the increase of $\Lambda_s$.  In conclusion, in the present work, all the low-energy predictions of $|m_{ee}|$ for $\tan\beta = 30$ and $58$ with $\Lambda_s = 1\ TeV$, $7\ TeV$ and $14\ TeV$ are found to be consistent with the experimental upper bound $|m_{ee}| < (0.028 - 0.122)\ eV$ in both NO and IO scenarios.

\end{document}